\documentclass{revtex4}

\usepackage{graphicx}
\usepackage{amsmath}
\usepackage{hyperref}

\newcommand{\Br}[2]{\mathrm{Br}^{(\chi_{c#1})}_{{#2}}}
\newcommand{\fBr}[2]{\frac{\Br{#1}{#2}}{\Br{#1}{\gamma}}}
\newcommand{\Brb}[2]{\mathrm{Br}^{(\chi_{b#1})}_{{#2}}}
\newcommand{\fBrb}[2]{\frac{\Brb{#1}{#2}}{\Brb{#1}{\gamma}}}
\newcommand{\JP}{{J/\psi}}
\newcommand{\LL}{{\ell\ell}}

\newcommand{\GeV}{\mathrm{GeV}}
\newcommand{\A}[1]{\mathcal{A}^{(#1)}}
\begin{document}

\author{A.~V.~Luchinsky}
\title{Muon Pair Production in Radiative Decays of Heavy Quarkonia}
\begin{abstract}
  Muon pair production in exclusive radiative decays $\chi_{cJ}\to\JP\mu\mu$, $\chi_{bJ}\to\Upsilon\mu\mu$ are considered. We present numerical values of the branching fractions and distributions over various kinematical variables.
\end{abstract}
\affiliation{NRC  ``Kurchatov Institute''-IHEP, Protvino, Russia}
\maketitle


The study of heavy quarkonia, i.e. mesons build from heavy (with $m_Q\gg\Lambda_\mathrm{QCD}$) quark-antiquark pair is a very interesting task from both theoretical and experimental points of view. The theoretical model that is usually used for theoretical description of these particles is the Nonrelativistic Quantum Chromodynamic (NRQCD) \cite{Bodwin:1994jh}, that allows one to describe with pretty good accuracy the processes of charmonia (e.g. $\JP$ or $\chi_{cJ}$ or bottomonia (e.g. $\Upsilon$, $\chi_{bJ}$) production on hadronic colliders as well as various decays of these particles \cite{Aad:2011sp,Chatrchyan:2011kc,TheATLAScollaboration:2013bja,Butenschoen:2012qr,Butenschoen:2012px, Likhoded:2016zmk, Aaij:2016bqq}. It should be mentioned, however, that NRQCD predictions depend on a number of parameters (so called NRQCD matrix elements), whose numerical values are determined phenomenologically from analysis of available experimental data \cite{Likhoded:2014kfa,Abe:1997yz,Abulencia:2007bra,Chatrchyan:2012ub,LHCb:2012ac,Aaij:2013dja}.

In addition to mentioned above processes there are also reactions that can be considered in almost model independent way. Among such processes one can name, for example, lepton pair production in $\chi_{cJ}\to\JP\ell\ell$ and $\chi_{bJ}\to\Upsilon\ell\ell$ decays. In \cite{Faessler:1999de} it was shown that the branching fractions of these decays and distributions over the invariant mass of $(\ell\ell)$ pair can be calculated using very general assumptions on the basis of experimentally known branching fractions of the corresponding radiative decays $\chi_{cJ}\to\JP\gamma$ and $\chi_{bJ}\to\Upsilon\gamma$ (see also \cite{Eichten:1979ms,Brambilla:2004wf, Barnes:2005pb,Cao:2016xqo}). Currently only $\chi_{c1,2}\to\JP ee$ process was studied experimentally \cite{Ablikim:2017kia}. It could be interesting also to consider muon pair production in $\chi_{cJ}\to\JP\mu\mu$ and $\chi_{bJ}\to\Upsilon\mu\mu$ decays. In the current paper we consider theoretically these reactions.



\begin{figure}
  \centering
  \includegraphics[width=\textwidth]{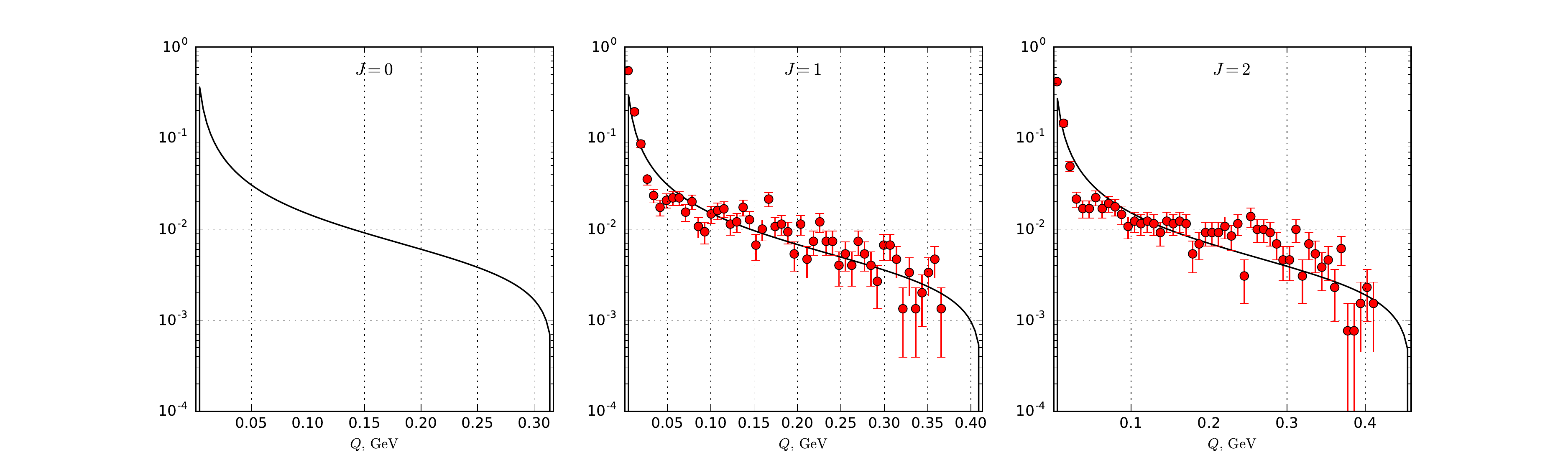}
  \caption{$Q$ distribution in comparison with BESSIII data \cite{Ablikim:2017kia, Zhang:PC}}
  \label{fig:hQEE}
\end{figure}

In the recent experimental paper \cite{Ablikim:2017kia} BESIII Collaboration analysed electron-positron pair production in $\chi_{c1,2}$-meson decays
\begin{align}
  \label{eq:decay}
  \chi_{cJ} &\to\JP e^+ e^-.
\end{align}
 In this article branching fractions of the named reactions and distributions over invariant mass of the lepton pair were presented. It is interesting to note, that these results can be obtained from the branching fractions of the radiative decays $\chi_{cJ}\to\JP\gamma$ using a very simple relation. In paper \cite{Faessler:1999de} for example, it was shown, that $q^2=m_{ee}^2$ distribution of the (\ref{eq:decay}) decay is equal to
 \begin{align}
   \label{eq:II}
   \frac{d\Br{J}{\LL}}{dq^2} &= 
                               \frac{\alpha}{3\pi q^2}\frac{\lambda(M_\chi,M_\psi,\sqrt{q^2})}{\lambda(M_\chi,M_\psi,0)} 
                               \left(1+\frac{2m_\ell^2}{q^2}\right)\sqrt{1-\frac{4m_\ell^2}{q^2}}
                               \Br{J}{\gamma},
 \end{align}
where $\Br{J}{\LL}$ and $\Br{J}{\gamma}$ are the branching fractions of $\chi_{cJ}\to\JP\LL$ and $\chi_{cJ}\to\JP\gamma$ decays respectively and
\begin{align}
  \lambda(M,m_1,m_2) &= \sqrt{1-\left(\frac{m_1+m_2}{M}\right)^2} \sqrt{1-\left(\frac{m_1-m_2}{M}\right)^2}
\end{align}
is the velocity of the final particle in $M\to m_1m_2$ decay. It should be noted that this result is almost model independent since it is based on the gauge invariance of $\chi_c\to\JP\gamma^*$ vertex. The only assumption that was made is that one can neglect the form factors' dependence on photon virtuality. This assumption seems pretty reasonable since according to energy conservation $q^2<(M_\chi-M_\psi)^2\sim 0.025\,\GeV^2$, that is much smaller than the typical hard scale $\sim M_\psi^2\sim 10\,\GeV^2$. It is clear from presented in Fig.~\ref{fig:hQEE} figures, that obtained by BESIII Collaboration experimental data are in good agreement with theoretical predictions (\ref{eq:II}). The same is true also for the integrated values of the branching fractions: theoretical predictions in comparison with BESIII results are
\begin{align}
 \fBr{0}{ee} &= 8.1\times 10^{-3},\qquad   \left(\fBr{0}{ee}\right)_\mathrm{exp} = (9.5\pm1.9\pm0.7)\times 10^{-3},\\
 \fBr{1}{ee} &= 8.6\times 10^{-3},\qquad   \left(\fBr{1}{ee}\right)_\mathrm{exp} = (10.1\pm0.3\pm0.5)\times 10^{-3},\\
 \fBr{2}{ee} &= 8.7\times 10^{-3},\qquad   \left(\fBr{2}{ee}\right)_\mathrm{exp} = (11.3\pm0.4\pm0.5)\times 10^{-3}.
\end{align}

\begin{figure}
  \centering
  \includegraphics[width=\textwidth]{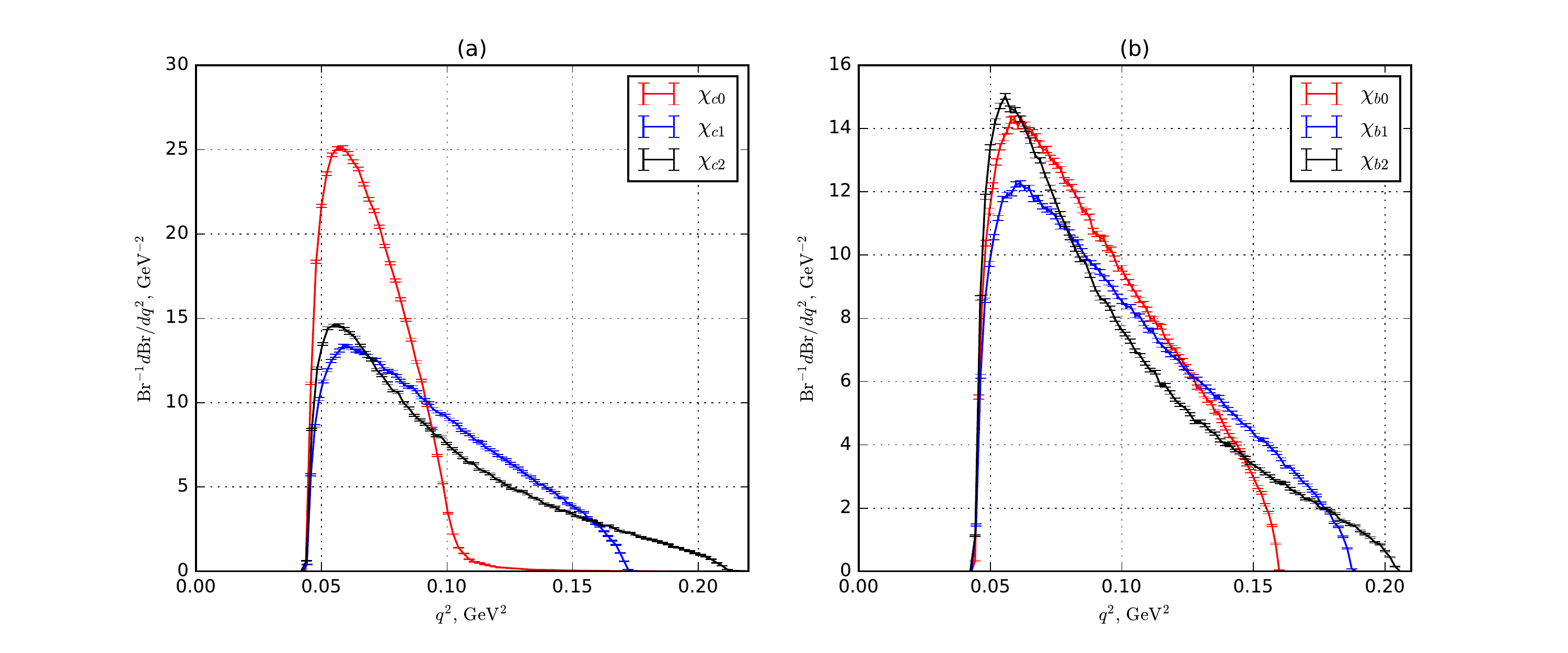}
  \caption{Normalized $Q^2$ distribution for $\chi_{cJ}\to J/\psi\mu\mu$ (left figure) and $\chi_{bJ}\to\Upsilon\mu\mu$ (right figure) decays}
  \label{fig:hQ2MM}
\end{figure}

It could be interesting to use described above formalism for muon pair production in $\chi_{cJ}\to\JP\mu\mu$ decays. It is clear that in this case we can also use (\ref{eq:decay}) relation to describe $q^2$ distributions, the corresponding results are shown in figure \ref{fig:hQ2MM}(a). One can easily see that the form of the distributions depends strongly on the spin of the initial particle and differs significantly from the presented in figure \ref{fig:hQEE} curves, but all these distributions are given by all universal relation (\ref{eq:II}), only the difference in participating particles' masses is important. As for the integrated branching fractions, we have
\begin{align}
\fBr{0}{\mu\mu} &= 2.2\times 10^{-4},
\quad \fBr{1}{\mu\mu}= 5.1 \times 10^{-4},
\quad\fBr{2}{\mu\mu}= 6.4\times 10^{-4}.
\end{align}
Using the same approach we have calculated also branching fractions of $\chi_{bJ}\to\Upsilon\mu\mu$ decays:
\
\begin{align}
  \fBrb{0}{\mu\mu} &= 4.7\times 10^{-4},
\quad \fBrb{1}{\mu\mu}= 5.7 \times 10^{-4},
\quad\fBrb{2}{\mu\mu}= 6.2\times 10^{-4}.
\end{align}
Distributions over leptons' invariant mass are shown in Figure \ref{fig:hQ2MM}(b).

\begin{figure}
  \centering
  \includegraphics[width=\textwidth]{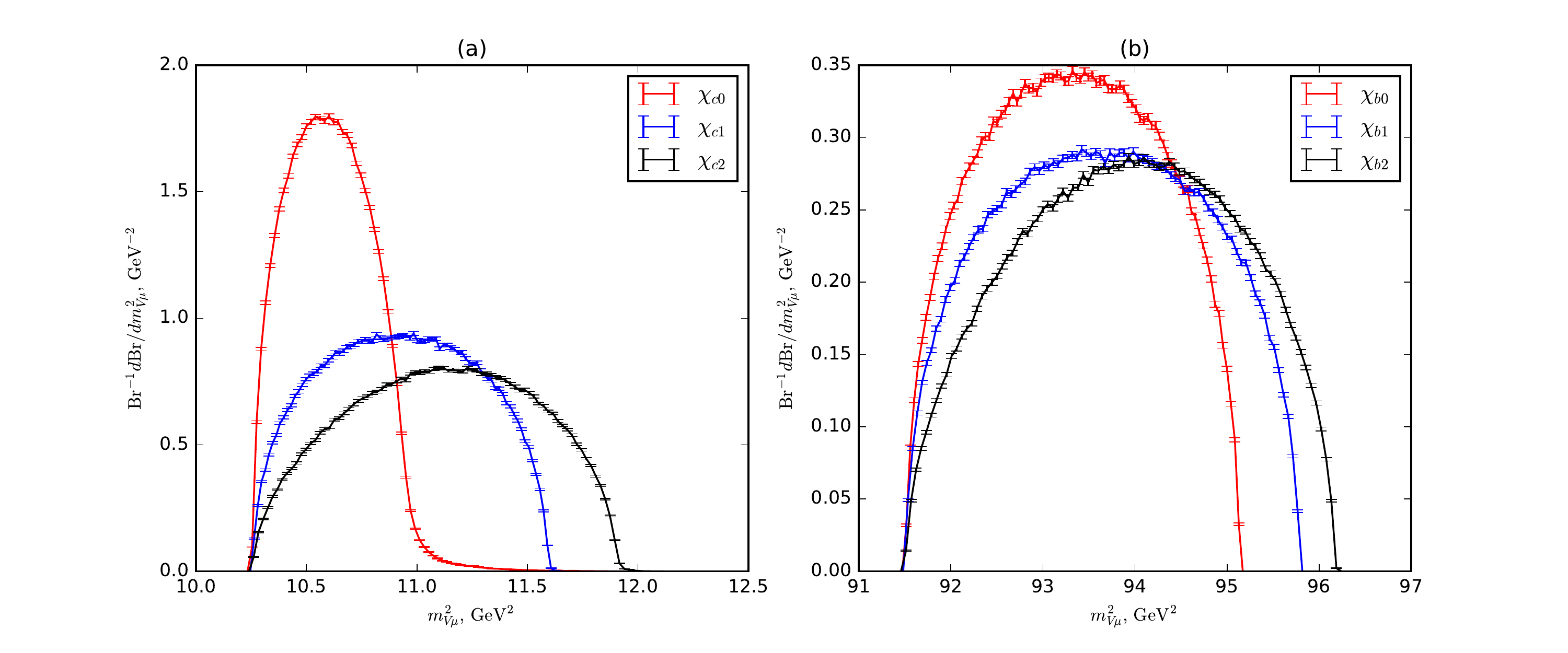}
  \caption{Normalized $m^2_{V\mu}$ distribution for $\chi_c$- and $\chi_b$ meson decays (left and right panels respectively)}
  \label{fig:hM2PsiM}
\end{figure}

If one wants to obtain other types of distributions more detailed information on the physics of the underlying process is required. In our work we use the following expressions for $P$-wave quarkonia  decay verticies \cite{Baranov:2011ib}:
\begin{align}
  \A{0} &= g_0 M_V \left( g_{\mu\nu} - \frac{p_\mu q_\nu}{(pq)}\right)\epsilon_V^\nu \epsilon^{(\gamma)}_\mu, \\
  \A{1} &= g_1 e_{\mu\nu\alpha\beta} q^\nu\epsilon_V^\alpha\epsilon_\chi^\beta \epsilon^{(\gamma)}_\mu,\\
  \A{2} &= \frac{g_2}{M_V} p^\mu \epsilon_V^\alpha \epsilon_{\chi_{c2}}^{\alpha\beta}\left[q_\mu\epsilon^{(\gamma)}_\beta-q_\beta\epsilon^{(\gamma)}_\mu\right],
\end{align}
where $\epsilon^{(\gamma)}$, $\epsilon_V$, and $\epsilon_\chi$ are polarization vectors of final photon, vector quarkonia and initial $\chi_Q$ meson respectively, while  dimensionless coupling constants $g_{0,1,2}$ can be determined from experimental values of the corresponding radiative decays.
Using these expressions it is easy to obtain presented in Fig.~\ref{fig:hM2PsiM} distributions over squared invariant of $\psi\mu$ pair.

\begin{figure}
  \centering
  \includegraphics[width=\textwidth]{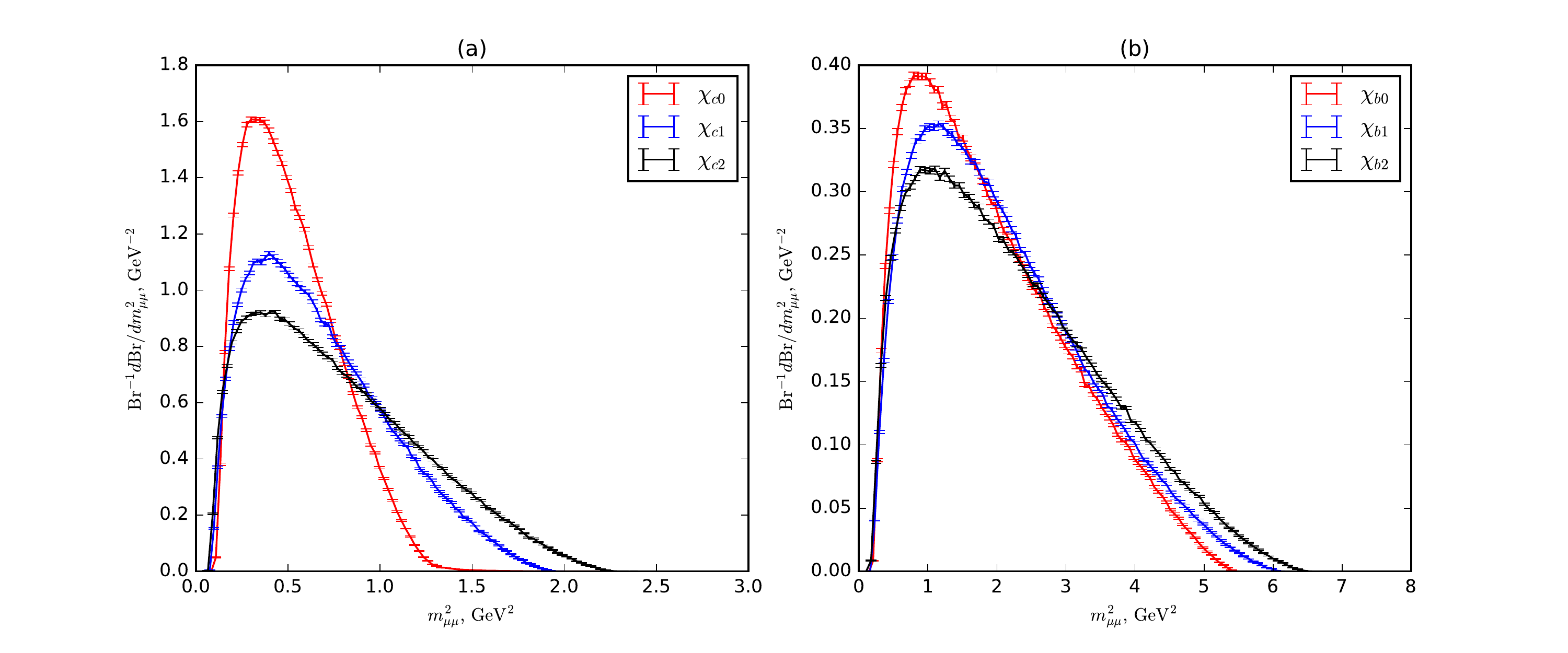}
  \caption{Normalized 
$m_{\mu^+\mu^+}^2$ distribution for $\chi_c$- and $\chi_b$ meson decays (left and right panels respectively)}
  \label{fig:hm2MM}
\end{figure}

Experimentally vector quakonium  is detected via its leptonic decay $V\to\mu^+\mu^-$. In order to obtain the information of the polarization picture of
\begin{align}
  \label{eq:dec2}
  \chi_{QJ} &\to V\mu^+\mu^- \to (\mu^+\mu^-)\mu^+\mu^-
\end{align}
it could be interesting to study also distributions over squared invariant mass of the same-signed final leptons. These distributions are shown in Fig.~\ref{fig:hm2MM}.

The author would like to thank I. Belyaev for useful discussions and J. Zhang for providing BESIII experimental data.


%

\end{document}